\documentclass[preprint,12pt]{revtex4-1}
\usepackage{graphicx}
\usepackage{dcolumn}
\usepackage{bm}
\RequirePackage{lineno}

\begin{document}


\title{Stratified chaos in a sand pile formation}

\author{Ate Poortinga}
 \email{ate.poortinga@wur.nl}
\affiliation{%
Soil and Land Management Group, Wageningen University, P.O. Box 47, 6700 AA Wageningen, The Netherlands
}%

\author{Jan G. Wesseling}
\affiliation{
Team Soil Physics and Land Use, Alterra - Wageningen University and Research Center, P.O.Box 47, 6700 AA Wageningen, The Netherlands
}%

\author{Coen J. Ritsema}
\affiliation{%
Soil and Land Management Group, Wageningen University, P.O. Box 47, 6700 AA Wageningen, The Netherlands
}%


\begin{abstract}
Sand pile formation is often used to describe stratified chaos in dynamic systems due to self-emergent and scale invariant behaviour. Cellular automata (Bak-Tang-Wiesenfeld model) are often used to describe chaotic behaviour, as simulating physical interactions between individual particles is computationally demanding. In this study, we use a state-of-the-art parallel implementation of the discrete element method on the graphical processing unit to simulate sand pile formation. Interactions between individual grains were simulated using a contact model in an Euler integration scheme. Results show non-linear self-emergent behaviour which is in good agreement with experimental results, theoretical work and self organized criticality (SOC) approaches. Moreover, it was found that the fully deterministic model, where the position and forces on every individual particle can be determined every iteration has a brown noise signal in the x and y direction, where the signal is the z direction is closer to a white noise spectrum.

\end{abstract}

\maketitle


\section{Introduction}
Non-linear dynamics in complex systems is an important topic of investigation in various scientific disciplines. Complex non-linear systems exhibit self-emergent behavior at macroscopic scale, driven by processes acting on the microscopic scale. Sand-pile formation is often used to describe this self-organizing behavior. When dry sand is poured on a surface, a conical pile will be formed with an angle of repose around 34 degrees. Additional particles result in an unstable state causing the structure to topple into a state where gravitational and frictional forces are in an equilibrium. Literature on the this topic is vast and the stratified chaos in sand pile formation has become a metaphor, driven by the Bak-Tang-Wiesenfeld (BTW) model \citep{Bak1987381, bak1993}, who laid the basis for  self-organized criticality (SOC). SOC is nowadays  widely used in different scientific disciplines such as economics \citep{Cont2000170}, neural science \citep{Buzsaki20041926} and earthquake research \citep{Bak20021785011}.

Pile (sand or rice) formation exhibits scale invariant behavior driven by small scale non-linear system properties \citep{PhysRevLett.65.1120,PhysRevLett.83.3952,PhysRevA.43.7091}. The system contains a critical attractor with events of various sizes that follow a power law. Like many natural systems, the temporal signal of sand-pile formation is characterized by a $1/f^{2}$ spectrum \citep{jensen19891}. Besides experimental studies \citep{Fret,PhysRevLett.65.1120,PhysRevLett.92.058702}, a variety of approaches, varying from cellular automata \citep{Bak1987381,bak1993} to numerical integration models \citep{jensen19891} have been conducted to describe the deterministic chaos in sand pile formation. Recently, it has become possible to describe the physical behavior of all single features at microscopic scale to study self-emergent behavior at the macroscopic scale due to the increasing computational power and efficient implementations of computational intensive tasks.  

Studies by e.g. \citet{Bell200577,Iglberger2010105,Longmore2013983,Stahl2011417} are examples of how particle based computations are integrated in a discrete element model (DEM). As interactions between bodies ($n > 2$) can not be solved analytically, a time integration scheme and contact model are used to describe the forces between particles. The behavior of granular matter can be simulated in a realistic manner by computing all forces acting on a particle and calculating their effects in three dimensions within a specific time step. This new generation of scale-invariant models can be applied to study stratified chaos in more detail (i.e. in combination with variation in gravitational forces, particle geometry and mass, frictional forces etc.). In this paper it is demonstrated that, besides visually appealing representations, these novel approaches contain features of stratified chaos that are in agreement with previous experimental and theoretical findings. This paper aims to compare experimental work with numerical simulations on a particle scale. 

\section{Materials and Methods}
A high-speed camera was used to record the formation of a sand-pile. For simulation purposes, we have used the framework of \citet{Longmore2013983}, an adoption of the method as described by \citet{Bell200577}, which concept was originally developed by \citet{Cundall197947}. The framework was written in C++, the OpenGL shading language GLSL and was rendered in OpenGL. Calculations were performed on a Graphic processing Unit (GPU) as this allows to perform the computations in parallel, thus significantly reducing the calculation time. The model runs were done on a regular desktop PC, containing two graphical cards (gtx 560ti) in a SLI configuration. In the simulations, we used spheropolygonal grains (a tetrahedral arrangement of four spherical particles), to ensure computations can be handled efficiently while maintaining static friction due to interlocking \citep{Pchel1,Longmore2013983} and preventing stick-slip behavior \cite{Bell200577}. During each iteration (Euler integration), the individual forces working on a particle are updated and summed to a total force (for the grain). The total force $F$ (N) for particle p in particle collection P can be calculated by the particle mass $m$ (kg) and gravitational acceleration $g$ ($m.s^{-2}$) :

\begin{equation}
F_{p}  =\sum \limits_{i \in P - {p}} F_{i} +mg 
\end{equation}

The total force working on a grain is calculated by summing the forces acting on the connected particles. These forces are split into the total normal force ($F^{g}$) (N) and the torque ($T^{g}$)(N$\cdot$m)

\begin{equation}
F^{g} = \sum\limits_{i=1}^{N_{p}} F_{i}
\end{equation}
\begin{equation}
T^{g} = \sum\limits_{i=1}^{N_{p}} r_{i} \times F_{i}
\end{equation}

where $N_{p}$ is the number of connected particles and $r_{i}$ represents the relative vector from the center of the granule to its child particle $i$. With Newton's second law and the particle mass the acceleration at time $t$ can be calculated. 

The contact forces of two colliding particles $p_i$ and $p_j$ is divided into the normal force and the tangential component:
\begin{equation}
 \vec{F}_{ij} = \vec{F}{_{ij}^n} + \vec{F}{_{ij}^t} 
\end{equation}
To simplify the contact detection and force calculation, an area of overlap $\xi$ ($m^2$) between two particles is defined. In case of spherical particles, the normal vector ($\vec{N}$) and the overlap area can be found by

\begin{equation}
\vec{N}_{ij} = \frac{\vec{X}_j - \vec{X}_i}{\parallel\vec{X}_j - \vec{X}_i\parallel}
\end{equation}
and
\begin{equation}
\label{xiEq}
\xi_{ij} = max(0, R_i+ R_j - \parallel\vec{X}_i - \vec{X}_j\parallel)
\end{equation}

where $R_i$ is radius and $\vec{X}_i$ is the center of particle i.

It should be noted that eq. \ref{xiEq} defines $\xi_{ij}$ rather as a mutual compression or deformation than as an overlap, where the time interval of the model should be small enough to prevent soft sphere behavior. Using the normal vector $\vec{N}$, the relative velocity $\dot{\xi}_{ij}$ of the compression can be calculated as

\begin{equation}
\dot{\xi}_{ij} = (\vec{v}_j - \vec{v}_i) \cdot \vec{N}_{ij}  
\end{equation}

The most basic formulations of the tangential and normal forces, incorporating the dissipative and friction terms are:

\begin{equation}
\vec{F}{_{ij}^n} = (-k_\alpha\dot{\xi}{_{ij}^\alpha}\xi_{ij} - k_\beta\xi{_{ij}^\beta)\vec{n}_{ij}}
\end{equation}

\begin{equation}
\vec{F}{_{ij}^t} = -min(\mu_s  \cdot \parallel\vec{F}{_{ij}^n}\parallel,k_t  \cdot \parallel\vec{V}_{ij}\parallel) \frac{\vec{V}_{ij}}{\parallel\vec{V}_{ij} \parallel }
\end{equation}

Where $\mu_s$ (N) represents friction. The viscous damping coefficient $k_\alpha$ is calculated from the coefficient of normal restitution ($e_n$), the reduced mass ($m_{eff}$) and the the time-step ($\Delta t$). A dimensionless coefficient of 0.02 was added to produce rapid damping:

\begin{equation}
k_\alpha = \frac{-2 \cdot m_{eff}  \cdot log( e_n )}{ (\Delta t \cdot 0.02) }
\end{equation}

where $m_{eff}$ is obtained from the mass of particles $m_1$ and $m_2$:
\begin{equation}
\frac{1}{m_{eff}} = \frac{1}{m_1}+\frac{1}{m2}
\end{equation}

The stiffness coefficient $k_\beta$ was calculated with Youngs parameter $E_{eff}$ and particle diameter $P_d$:

\begin{equation}
kr =  \frac{4}{3} \cdot ( E_{eff}) \cdot \sqrt{ 0.25 \cdot P_d }
\end{equation}

\begin{figure}[h!]
    \includegraphics[width=0.4\textwidth]{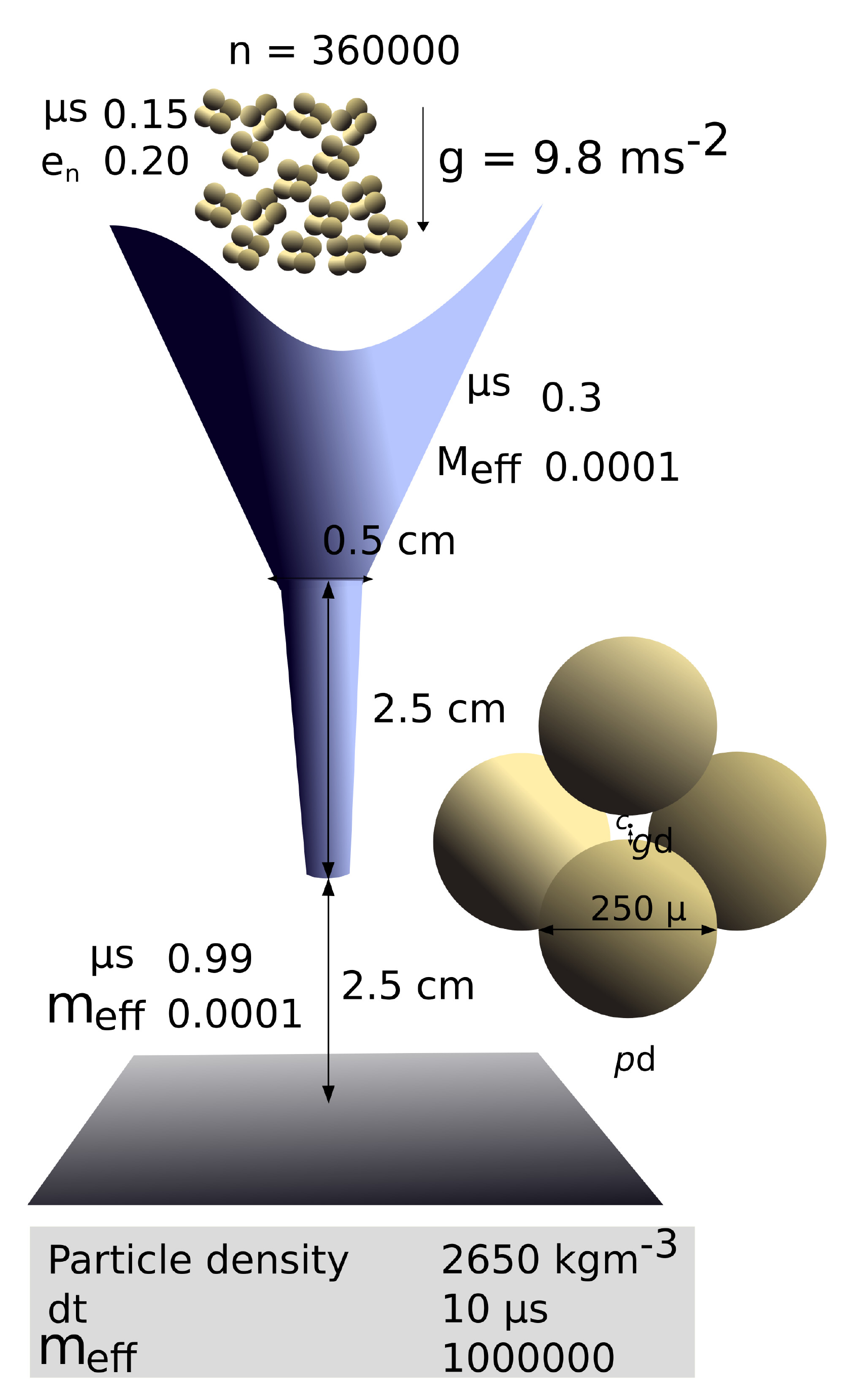}
	\caption{A schematic overview of the dimensions and parameter settings used in the model simulation and a representation of the spheropolygonal grains and cumulative distribution to center of mass. }
	\label{img:method}
\end{figure}

The parameters used in the simulation are shown in Fig. \ref{img:method}. Different coefficients for friction and restitution were used for the floor, funnel and particles. A small variation in distance to the center of mass of the spheropolygonal grains was included to emulate a variation in particle characteristics. Particle size and mass were kept constant for all grains, as a variation would add an extra layer of complexity to the study.

One of the problems in time integration is finding the optimal time step size, as time integration is a trade-off between computational efficiency and a physically correct representation of the process. A time-step of 10 $\mu$s was found to give reliable results for a particle diameter of $250 \mu$. However, in the current framework, data on the GPU can only be rendered directly to the screen (for visual interpretation) and not directly be read from the GPU memory. In order to obtain the data, a memory transfer from the GPU to the CPU is required, which is a computational very demanding task. After testing different sampling resolutions on relatively small datasets, a sampling resolution of 100 $\mu$s  was found to give reliable results.

\section{Results and Discussion}

\begin{figure}[h!]
    \includegraphics[width=1.0\textwidth]{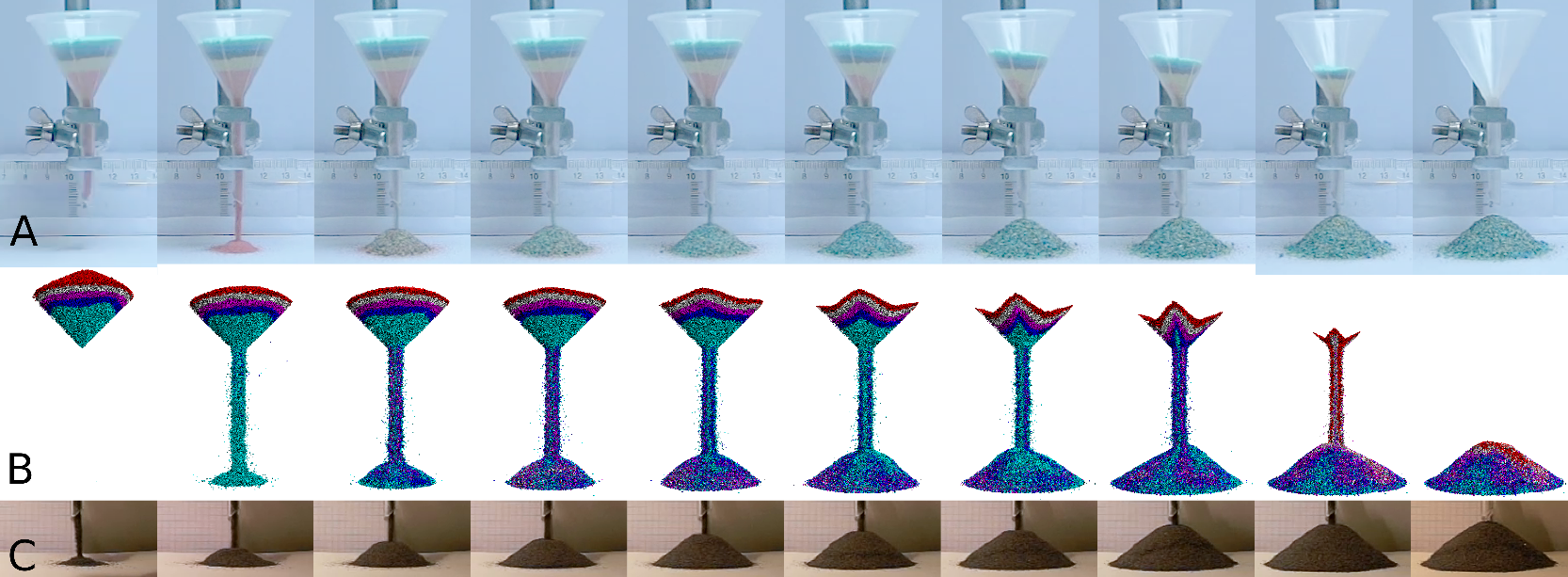}
	\caption{A visual comparison between an experiment with a layer of different colors (A), the model run with a layer of different colors (B) and an experiment with black sand focused on the avalanches of the sand-pile (C).}
	\label{img:2}
\end{figure}

In both laboratory and simulation experiments different layers of colored grains were added to the funnel for visual comparison with the experimental results (Fig. \ref{img:2} top and middle). The experimental results as well as the model results show a rapid mixing between the different colors. The different layers mix in the middle of the funnel, while maintaining the layered structure on the sides in the funnel. The sand flow contains different colors that form a cone with an angle of repose of approximately 34 degrees. The particles added to the top of the sand-pile result in a non-equilibrium situation where the angle of repose exceeds the 34 degrees, resulting in an avalanche where the sand from the top flows over the surface of the sand-pile. This is clearly visible in both the experimental and the modeled results (Fig. \ref{img:2} middle and bottom)

\begin{figure}[h!]
    \includegraphics[width=0.6\textwidth]{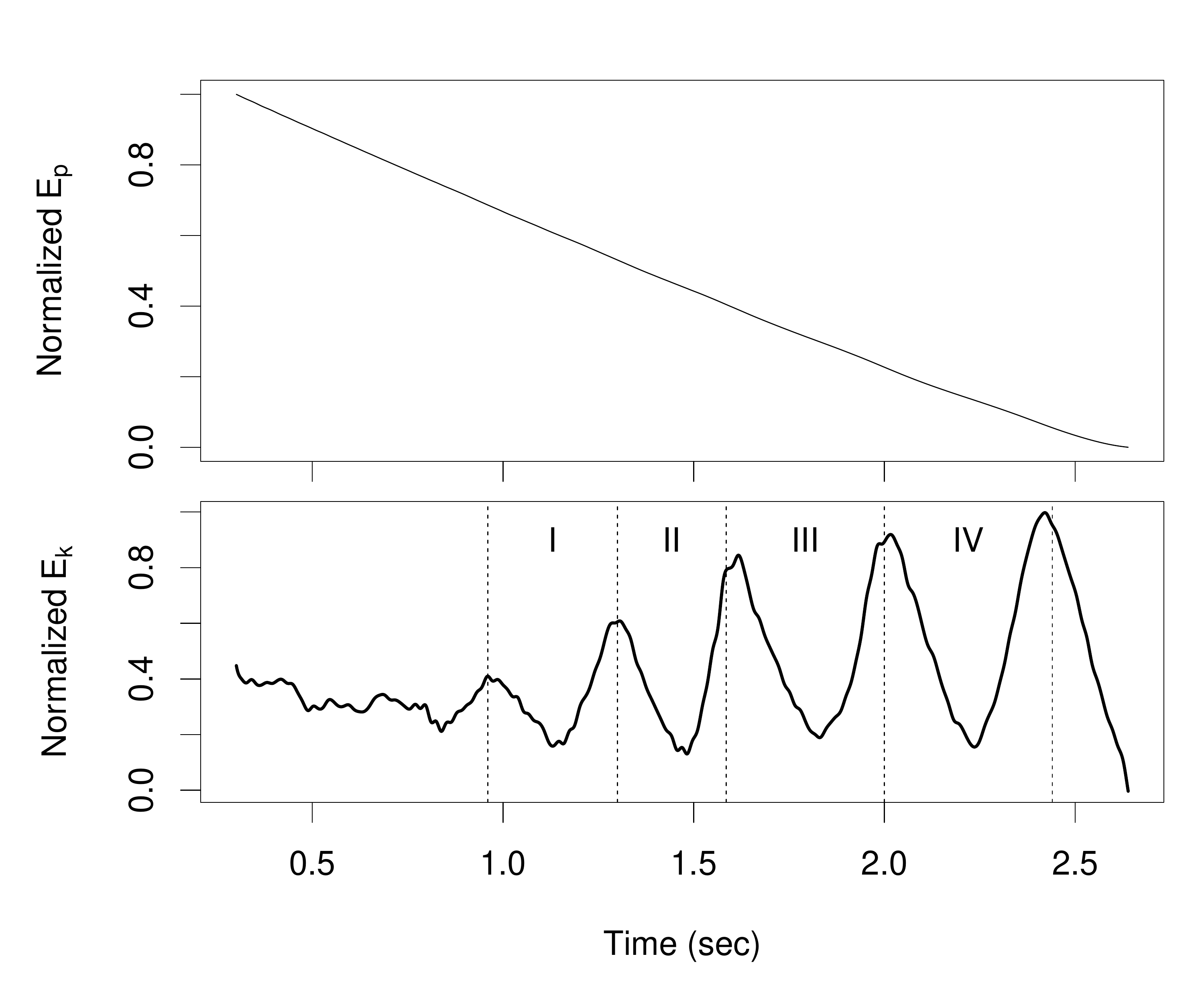}
	\caption{The normalized potential energy of the entire system (left), indicating a constant throughput of particles and the normalized kinetic energy for the sand-pile (right), with periods of energy dissipation and absorption. The kinetic energy release was divided into four different periods (I II, III, IV) for further analysis.}
	\label{img:3}
\end{figure}

The constant grain throughput of the funnel leads to a linear decrease in potential energy (Fig. \ref{img:3}). The entropy of the system diverges through intermittent dissipation of kinetic energy in the sand-pile (Fig. \ref{img:3}). With gravitational forces exceeding the frictional ones, the sand-pile shows relaxation oscillations where kinetic energy is dissipated by avalanches. Considering the constant input of energy into the system, an increase in kinetic energy implies energy dissipation, whereas a decrease in kinetic energy represents the build up of energy. Previous studies \citep{PhysRevLett.65.1120,Fret} used the fluctuation in the mass of a sand-pile to study the energy distribution in a sand-pile formation. We are able to use the kinetic energy directly to study the fluctuations in energy distribution.

\begin{figure}[h!]
    \includegraphics[width=0.45\textwidth]{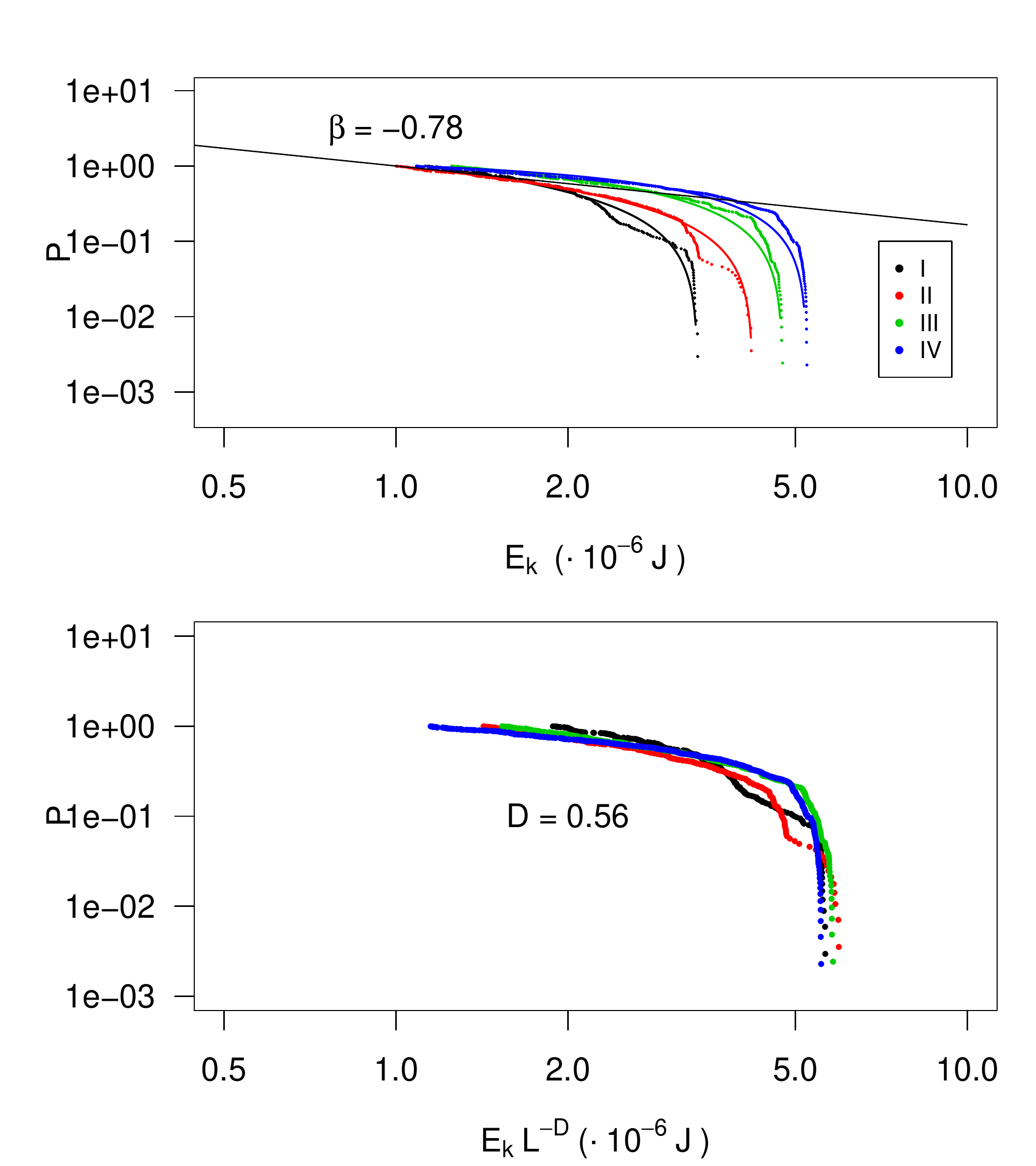}
	\caption{The cumulative probability distribution of avalanches for four different periods (see Fig. \ref{img:3}). The data show that periods of energy dissipation follow a power law and finite-size scaling for the different periods. The data scales following the parameter $D = 0.56$. }
	\label{img:4}
\end{figure}

The cumulative frequency-magnitude probability of the kinetic energy shows power law behaviour:
\begin{equation}
P(E_{k})= {E_{k}}^{\beta}
\end{equation}
with $\beta = -0.78$. It was found that the sand-pile has a finite-size scaling behaviour for the growing sand-pile (for periods I, II, III and IV, Fig. \ref{img:4}). The lines fitted through the data-points include the upper and lower cut-off values in an $upper\: truncated \: Pareto\: distribution$ (equation 12) \citep{hergarten2002self}. We have found $\beta  = $ 0.21, 0.16, -0.29 and -0.81 for period I, II, III and IV respectively. The cumulative frequency-magnitude probability $E_{k}$ for energy dissipation for a pile with  potential energy $E_{p}$ can be given by equation \ref{energy} \citep{barabas}, with $f(x)$ constant up to some value and $D = 0.56$ (Fig. \ref{img:4} bottom)

\begin{equation}
\label{energy}
 P(E_{k}) = \frac{(\frac{E_{k}}{E_{k_{min}}})^{-\beta} - (\frac{E_{k_{max}}}{E_{k_{min}}})^{-\beta}}{1 - ( \frac{E_{k_{max}}}{{E_{k_{min}}}} )^{-\beta}} \:  for  \: E_{k_{min}} < E_{k} < E_{k_{max} }  
\end{equation}

\begin{equation}
 P(E_{k},E_{p}) = E_{k}^{-\beta} f \frac{E_{k}}{E_{p}^{D}} 
\end{equation}

\begin{figure}[h!]
    \includegraphics[width=0.9\textwidth]{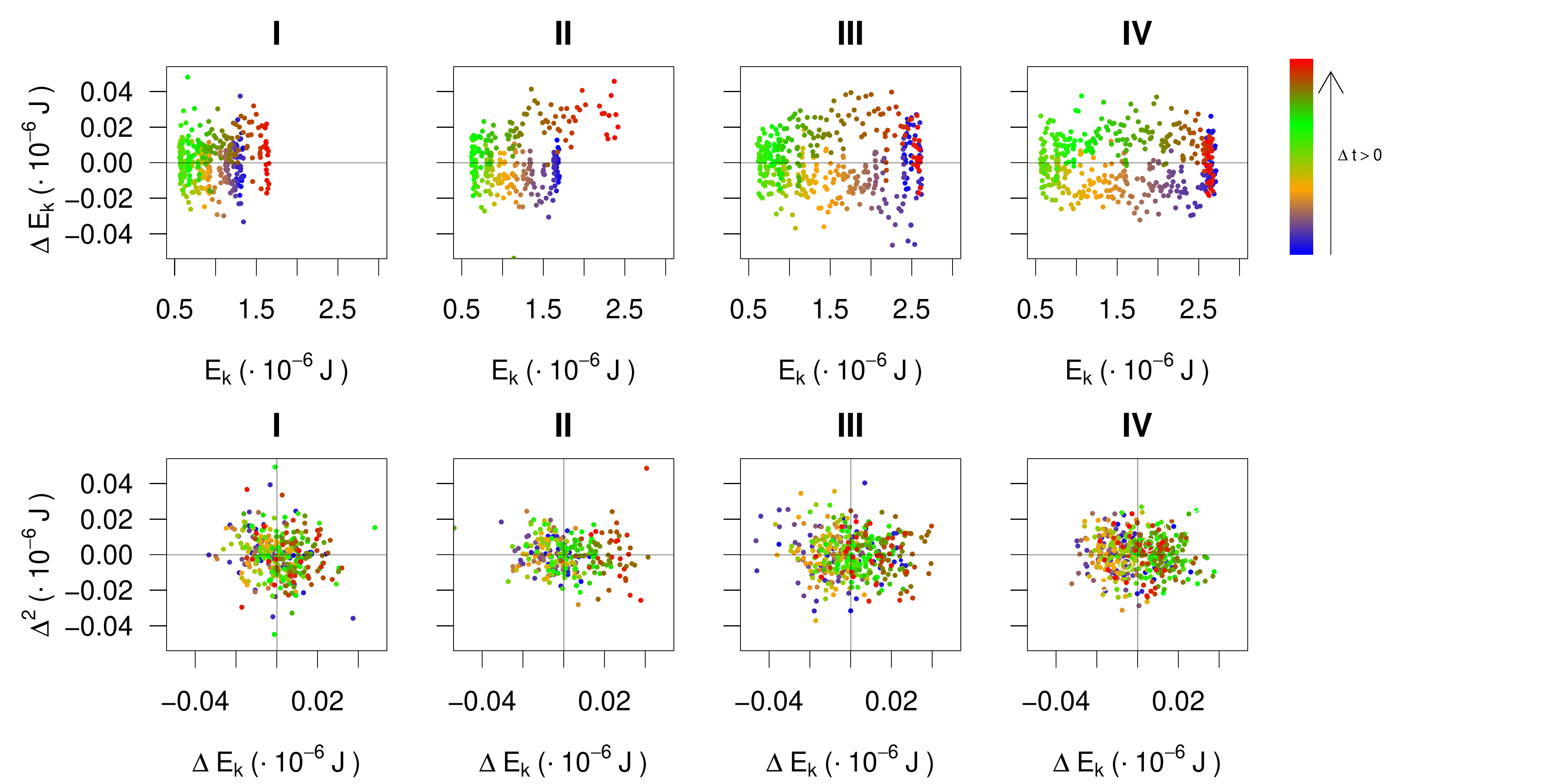}
	\caption{The energy distribution for periods I, II, III and IV with $E_{k}$ versus $\delta E_{k}$ (top) and $\delta E_{k}$ versus $\delta^{2} E_{k}$ bottom. Scale invariance is governed by the build-up of energy, where $\delta E_{k}$ remains within the same range for the different periods.}
	\label{img:5}
\end{figure}

Scale invariance is governed by the build-up of energy in the sand-pile since the $\delta E_{k}$ has the same order of magnitude for the different time periods (I,II,III and IV, Fig. \ref{img:5}). Fig. \ref{img:5} (top) clearly shows that energy build-up (and thus dissipation) increases with the growing sand-pile due to consecutive steps of energy build-up through time. This implies that finite sized scaling is governed by consequential build-up and dissipation of $\delta E_{k}$ and not by differences in $\delta E_{k}$. Due to the laws of entropy, the system is attracted towards the low energy state, an equilibrium situation between friction and gravity. The $\delta E_{k}$ and derivative $\delta^{2}E_{k}$ (Fig. \ref{img:5} bottom) have no scale dependency and show a chaotic behavior within the specific domain.

\begin{figure}[h!]
    \includegraphics[width=0.75\textwidth]{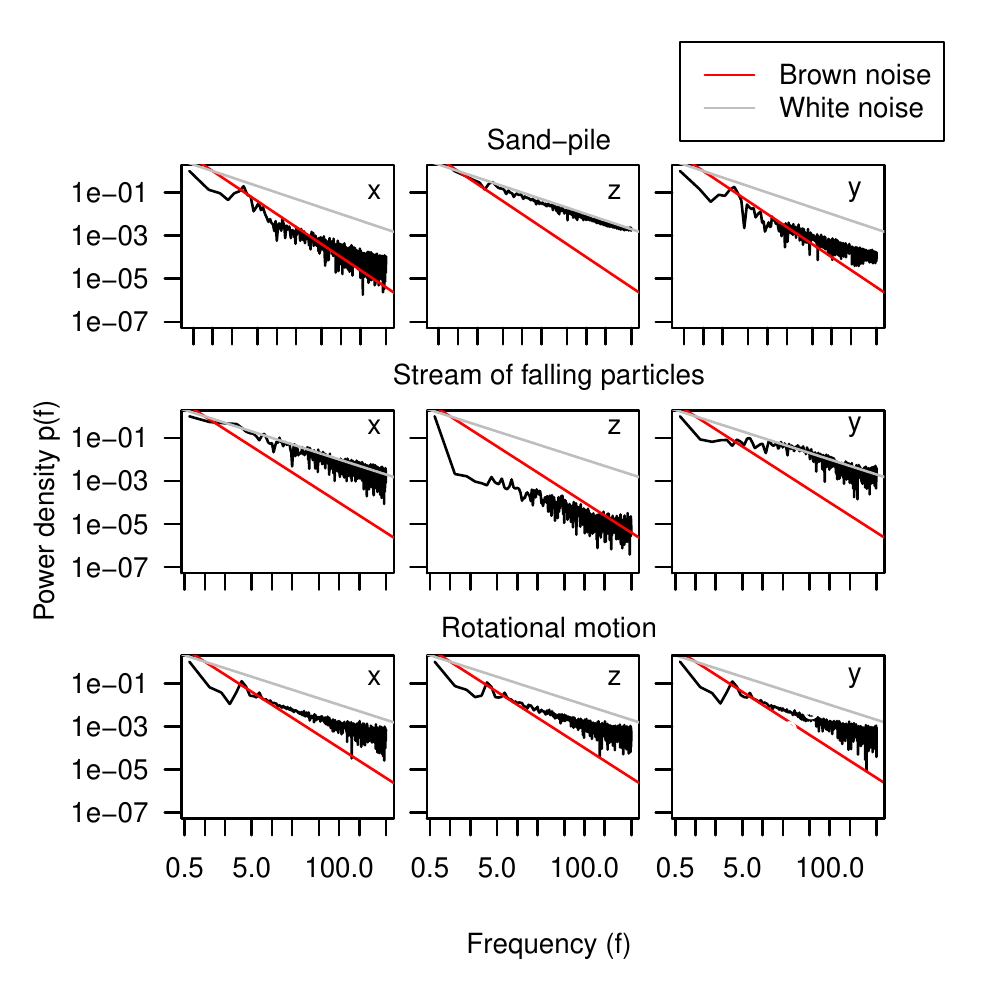}
	\caption{Power spectrum of the sand-pile (top), stream of particles from the funnel (middle) and rotational motion (bottom).}
	\label{img:6}
\end{figure}

\citet{Bak1987381} used the BTW model to explain the ubiquitous $1/f$ noise signal, found in many natural systems, whereas an experimental study \citep{jaeger1989relaxation} found no $1/f$ noise signal. Later studies confirmed that this was in fact a brown $1/f^{2}$ noise spectrum \citep{jensen19891,kertesz1990noise}. Similar results were found in this study, the power spectrum of the kinetic energy of the sand pile in x and y direction has a brown noise signal ($1/f^{2}$) (Fig. \ref{img:6}). However, the kinetic energy in the vertical (z) direction is closer to a white noise spectrum ($1/f$). This means that the frequencies of the distribution in the z direction have the same amplitude, whereas there is a dominance to low frequencies in the x and y direction. When boundary conditions are included (i.e. interaction with the floor) this white noise changes to a brown noise spectrum. The stream of particles from the funnel have a white noise spectrum in the x, y and z direction. The rotational motion of the grains have a pink noise spectrum ($1/f^{\beta}$) in the x, z and y direction .

Previous experimental work used the falloff (of mass) from a sand-pile of every time-step to study the behavior of avalanches. In this study the frequency distribution of energy dissipation was used in the same manner. However, an avalanche is the build-up and dissipation of energy over a larger number of consecutive time-steps and was therefore not used in this study. Fig. \ref{img:3} shows for example five events of large energy dissipation which can be labelled as an avalanche. Furthermore, it should be noted that whereas physical considerations form the core of the current approach, the correctness of the output is still dependent on the parameter settings in the contact model and the numerical integration scheme. 

Self organization and chaos are important characteristics in many natural systems. Though often applied, SOC approaches are controversial in explaining non-linear dynamics in complex system because they lack any physical basis. Computational limitations remain a constraint for the current framework (in the order of days), however, with the ever increasing computational power and more efficient implementations, these physical based approaches can find their way in a variety of scientific disciplines in the near future. Specifically. studies involving the physics particle movement could benefit from this approach, deployed in this study.

\section{Conclusion}

We have shown that a simple set of rules, defined here as gravity and particle interactions in a contact model, results in non-linear self-emergent behavior which is in good agreement with experimental results, theoretical work and SOC approaches. The model is fully deterministic, i.e. the position and forces on every individual particle can be determined every iteration, while showing complex non-linear self-organizing behavior. Furthermore, 
it offers the possibility to predict occurrence and avalanche behaviour in growing sand piles.

\section{Acknowledgement}
We would like to offer special thanks to the Department of Computer Science, University of Cape Town, for sharing their source code, with specific gratitude to Juan-Pierre Longmore and Craig Leach. We thank Paul Torfs for his comments on the model simulations.

\newpage

\bibliography{ref}	



\end{document}